\documentclass[useAMS,usenatbib]{mn2e}

\usepackage{graphicx}

\title[Accurate laboratory UV wavelengths]{Accurate laboratory ultraviolet wavelengths for quasar absorption-line constraints on varying fundamental constants}

\author[Aldenius, Johansson \& Murphy]{M.~Aldenius$^{1}$\thanks{E-mail: maria@astro.lu.se}, S.~Johansson$^{1}$ and M.~T.~Murphy$^{2}$\\
$^{1}$Atomic Astrophysics, Lund Observatory, Lund University, Box 43, SE-221 00 Lund, Sweden\\
$^{2}$Institute of Astronomy, University of Cambridge, Madingley Road, Cambridge CB3 0HA, UK}

\begin{document}

\date{Accepted 2006 April 26}

\pagerange{\pageref{firstpage}--\pageref{lastpage}} \pubyear{}

\maketitle

\label{firstpage}

\begin{abstract}
The most precise method of investigating possible space-time variations of the fine-structure constant, $\alpha\equiv (1/\hbar c)(e^{2}/4\pi \epsilon_{0})$, using high-redshift quasar absorption lines is the many-multiplet (MM) method. For reliable results this method requires very accurate relative laboratory wavelengths for a number of UV resonance transitions from several different ionic species. For this purpose laboratory wavelengths and wavenumbers of 23 UV lines from Mg{\sc \,i}, Mg{\sc \,ii}, Ti{\sc \,ii}, Cr{\sc \,ii}, Mn{\sc \,ii}, Fe{\sc \,ii} and Zn{\sc \,ii} have been measured using high-resolution Fourier Transform (FT) spectrometry. The spectra of the different ions (except for one Fe{\sc \,ii} line, one Mg{\sc \,i} line and the Ti{\sc \,ii} lines) are all measured simultaneously in the same FT spectrometry recording by using a composite hollow cathode as a light source. This decreases the relative uncertainties of all the wavelengths. In addition to any measurement uncertainty, the wavelength uncertainty is determined by that of the Ar{\sc \,ii} calibration lines, by possible pressure shifts and by illumination effects. The absolute wavenumbers have uncertainties of typically $\pm 0.001$ to $\pm 0.002$\,cm$^{-1}$ ($\Delta \lambda \approx$ 0.06 to 0.1\,m\AA\ at 2500\,\AA), while the relative wavenumbers for strong, symmetric lines in the same spectral recording have uncertainties of $\pm 0.0005$\,cm$^{-1}$ ($\Delta \lambda \approx$ 0.03\,m\AA\ at 2500\,\AA ) or better, depending mostly on uncertainties in the line fitting procedure. This high relative precision greatly reduces the potential for systematic effects in the MM method, while the new Ti{\sc \,ii} measurements now allow these transitions to be used in MM analyses.
\end{abstract}



\section{Introduction}\label{introduction}

Ultraviolet (UV) transitions of metallic ions, particularly those involving the ground or low-lying states, are important probes of the interstellar medium (ISM), providing direct information about the abundances and kinematics of gas clouds along the lines-of-sight toward bright stars. The advent of space-borne telescopes equipped with high-resolution UV spectrographs, such as the Goddard High Resolution Spectrograph (GHRS), the Space Telescope Imaging Spectrograph (STIS) aboard the {\it Hubble Space Telescope} and the {\it Far Ultraviolet Spectroscopic Explorer} (FUSE), focused attention on the (im)precision of most UV resonance laboratory wavelengths \citep[e.g.][]{MortonD_91a}. Typical laboratory wavelength uncertainties were $\sim$2\,m\AA\ \citep{MortonD_91a} while the resolving powers of the astronomical spectrographs were as high as $R \sim 50000$ (25\,m\AA\ at 1250\,\AA). Thus, the analysis technique and physical information recoverable from the astronomical spectra were often compromised \citep[e.g.][]{LinskyJ_95a}.

The same UV transitions are observed in the optical region in the study of highly redshifted absorption-lines arising in galaxies and the intergalactic medium along the sight-lines to distant quasars (QSOs). The era of 8--10-m class ground-based telescopes with stable high-resolution spectrographs (e.g.~Keck/HIRES, VLT/UVES, Subaru/HDS) has seen a dramatic improvement in the quality and quantity of spectra covering a multitude of metal-line transitions, particularly in studies of damped Lyman-$\alpha$ systems \citep[e.g.][]{LuL_96a,ProchaskaJ_03c}. While it is clearly important for these general Galactic and extragalactic applications to have accurate and precise laboratory wavelengths, recent developments in the study of cosmological variations in the fundamental constants has demanded an order of magnitude improvement in the laboratory wavelengths of several ionized metallic transitions.

The universality and constancy of the laws of nature rely on the space-time invariance of fundamental constants, such as the fine-structure constant, $\alpha\equiv (1/\hbar c)(e^{2}/4\pi \epsilon_{0})$. Tests of the constancy of $\alpha$ therefore probe deviations from the standard model of particle physics. QSO absorption lines are a particularly attractive probe, constraining $\alpha$-variation over large spatial and temporal scales. Initial studies (e.g.~\citealt*{BahcallJ_67b}; \citealt{VarshalovichD_94a,CowieL_95a}) focused on comparison of the fine-structure splitting in alkali-doublets (ADs) of, for example, Si{\sc \,iv}, in QSO and laboratory spectra but, as demonstrated by \citet{MurphyM_01c}, the dominant error in these studies arose from laboratory wavelength uncertainties.

Considerable recent interest has focused on the many-multiplet (MM) method introduced by \citet*{DzubaV_99b,DzubaV_99a} and \citet{WebbJ_99a}. The MM method is a generalisation of the AD method, constraining changes in $\alpha$ by utilising many observed transitions from different multiplets and different ions associated with each QSO absorption system. It holds many important advantages over the AD method, including an effective order-of-magnitude precision gain stemming from the large differences in the sensitivity of light (e.g.~Mg, Si, Al) and heavy (e.g.~Fe, Zn, Cr) ions to varying $\alpha$. The MM method, applied to Keck/HIRES QSO absorption spectra, has yielded very surprising results, with the first tentative evidence for a varying $\alpha$ by \citet{WebbJ_99a} becoming stronger with successively larger samples (\citealt{MurphyM_01a,WebbJ_01a}; \citealt*{MurphyM_03a}; \citealt{MurphyM_04a}). Although these results have proven robust to many astrophysical and instrumental systematic errors \citep{MurphyM_01b,MurphyM_03a}, more recent studies of much smaller samples of VLT/UVES spectra suggest null results \citep{ChandH_04a,SrianandR_04a,QuastR_04a,LevshakovS_05a,ChandH_06a}. The discrepancy between the Keck/HIRES and VLT/UVES results is yet to be resolved.

The precision in the relative change in $\alpha$ achieved by the above MM analyses is $\delta(\Delta\alpha/\alpha) \sim 10^{-6}$ which corresponds to a velocity precision of $\delta v \sim 20{\rm \,m\,s}^{-1}$, or $\delta\lambda \sim 2\times 10^{-4}$\,\AA\ at 2500\,\AA\ -- a typical rest-frame wavelength for the transitions used in MM analyses. This is an order of magnitude more precise than most lines in \citet{MortonD_91a}. Significant recent effort has therefore been invested into improving the laboratory wavelengths of transitions which can be used in MM analyses.

High precision laboratory measurements of UV lines from Mg{\sc \,i}, Mg{\sc \,ii}, Cr{\sc \,ii}, Zn{\sc \,ii} and Ni{\sc \,ii} have previously been reported by \citet{Pick98,Pick00}. In this paper we have remeasured these lines (except the Ni{\sc \,ii} lines, which are beyond the range of the Lund Fourier Transform spectrometer) for confirmation. We have also included another set of lines from Mg{\sc \,i}, Ti{\sc \,ii}, Mn{\sc \,ii} and Fe{\sc \,ii}, which are all visible in QSO absorption spectra. In total we present high precision laboratory wavelength measurements of 23 UV lines from seven ionic species. The uncertainty in the laboratory wavelengths depends both on the measurement precision and the wavenumber calibration. To reduce the uncertainty in relative wavelengths, the spectra of all ionic species were produced in the same hollow cathode light source, using a composite cathode. This made it possible for the different spectra to be recorded simultaneously and the relative wavelengths are thus less dependent on the calibration uncertainty. Due to a combination of range in relative line intensities, a line blend, and detector and instrument response at different wavelengths, it was necessary to acquire three different spectra for this analysis. The wavelengths for all lines, apart from one weak Fe{\sc \,ii} line, are determined with the required uncertainty of $\pm 2\times 10^{-4}$\,\AA\ or better. For strong symmetric lines measured in the same recording the relative uncertainties are estimated to be $\pm 0.3\times 10^{-4}$\,\AA\ or better. Confidence in this level of uncertainty was ensured by investigations of calibration effects as well as isotope structure and hyperfine structure of the lines.


\section{Experimental method}\label{experiment}

Spectra of Mg{\sc \,i}, Mg{\sc \,ii}, Ti{\sc \,ii}, Cr{\sc \,ii}, Mn{\sc \,ii}, Fe{\sc \,ii} and Zn{\sc \,ii} were produced in a water cooled hollow cathode discharge lamp (HCL). The composite cathode used was made of pure Fe with small pieces of Mg, Ti, Cr, Mn and Zn placed in it. The cathode had an inner diameter of 7.0 mm and a length of 50 mm. A mixture of Ne and Ar was used as carrier gas. Ne generally produces higher signal-to-noise ratios while Ar is used for wavelength calibration. The mixture of about 1/3\,Ne and 2/3\,Ar gave a stable light source with good signal-to-noise ratios (SNR$>$100) for most lines used in this study. The light source was run with a current between 0.6 and 0.7\,A and a pressure between 0.9 and 1.0\,torr. The main excitation mechanism in the hollow cathode is electron collisions, but charge transfer with the carrier gas ions may in certain cases contribute to the production and excitation of singly ionized atoms \citep{Johan78}. Spectra were also recorded with a pure Fe cathode since the Fe{\sc \,ii} $\lambda$2586\,\AA\ line was blended with a Mn{\sc \,ii} line in the spectrum from the composite cathode. The Mn{\sc \,ii} transition in question is not one of interest in QSO absorption-line work -- it is between two excited states rather than between the ground (or very low-lying) state and an excited state.

The wavenumbers were measured with the Lund FT500 UV Fourier Transform (FT) spectrometer, which is optimized in the wavelength range of 2000--7000\,\AA. Spectra of two overlapping wavenumber regions were acquired with different photomultiplier detectors, one between 20000 and 40000\,cm$^{-1}$ and another between 25000 and 50000\,cm$^{-1}$. It was not possible to include all lines of interest in one recording because of the limitations in wavenumber range for the photomultipliers and the FT spectrometer. The Ti{\sc \,ii} lines are lower in wavenumber than the rest of the lines and were therefore measured in a separate spectral acquisition. This recording also included the Ar{\sc \,ii} calibration lines. The resolution of the measurements was 0.06\,cm$^{-1}$ for the measured spectrum between 25000 and 50000\,cm$^{-1}$ and 0.05\,cm$^{-1}$ between 20000 and 40000\,cm$^{-1}$. This was sufficient to completely resolve the single lines. Twenty scans were co-added to achieve good signal-to-noise ratios (SNR$>100$) for most of the lines in this study. 

The noise in spectra observed with FT spectrometry, using the type of light source and detectors stated above, is predominantly photon noise. White noise in the interferogram is transformed into the spectral region as white noise, with a constant level throughout the spectrum. All spectral lines seen by the detector contribute to this noise level. It is therefore a disadvantage to record lines outside the region of interest. In order to band limit the spectra, reducing the noise, solar-blind photomultipliers (Hamamatsu R166) were used for the UV region. These have a long-wavelength cut-off at about 3000\,\AA . For the lower wavenumber region, UV to visible, a wider range photomultiplier (1P28) was used. The two types of photomultipliers overlap in wavenumber regions, giving a small region for transfer of calibration between them (see Section~\ref{calibration}).
 
The light source was placed close to the aperture of the spectrometer and focused with a lens. This was to increase the signal-to-noise ratio for lines close to 2000\,\AA , where air absorbs most light. To minimize turbulence and to further reduce the absorption from air the spectrometer was evacuated to a pressure of about 0.015 torr.

To ensure the lines are unaffected by self-absorption the quantity of each metal inserted in the cathode had to be varied and tested through a large number of recorded spectra. This was especially the case for the Zn{\sc \,ii} doublet. With just a trace of Zn the lines were strong but unaffected by self-absorption. For the Mg{\sc \,i} lines the procedure was similar, but the strong $\lambda$2852\,\AA\ line and the much weaker $\lambda$2026\,\AA\ line had to be measured in different recordings. The weak $\lambda$2026\,\AA\ line had to be measured with more Mg in the cathode, which gave a clearly self-reversed profile for the strong $\lambda$2852\,\AA\ line. The $\lambda$2852\,\AA\ line was therefore measured in the same recording as the Ti{\sc \,ii} lines. The lamp conditions and spectrometer resolution for the three recordings are listed in Table~\ref{exptable}.

To further investigate the effects of possible self-absorption and pressure shifts of the lines, recordings were also made with different currents and pressures in the light source. The effects were considered negligible and are discussed in Sections~\ref{calibration},~\ref{mnhyperfine} and~\ref{isotope}.

\begin{table*}
 \caption{Lamp conditions, Resolution and Correction factors}
 \begin{tabular}{cccccccc}
 \hline
 Recording & Wavenumber & Cathode & Carrier gas & Pressure & Current & Resolution & $k_{\mathrm{eff}}$\\
 number & range (cm$^{-1}$) & {} & {} & (torr) & (mA) & (cm$^{-1}$) & \\
 \hline
 I & 20538-41074 & Composite & Ar \& Ne & 0.9 & 700 & 0.05 & $(-2.12\pm 0.02)\cdot10^{-6}$\\
 II & 25277-50553 & Composite & Ar \& Ne & 1.0 & 600 & 0.06 & $(-2.23\pm 0.03)\cdot10^{-6}$\\
 III & 26857-53713 & Pure Fe & Ar & 1.0 & 700 & 0.04 & $(-2.54\pm 0.03)\cdot10^{-6}$\\
 \hline
 \end{tabular}
 \label{exptable}
\end{table*}


\section{Analysis}\label{analysis}

\subsection{Wavenumber calibration}\label{calibration}
Spectra observed by FT spectrometry have a linear wavenumber scale, whose accuracy derives from the control of the sampling of the interferogram by a single-mode helium-neon laser, whose frequency is stabilized to 5 parts in $10^{9}$. The accuracy is however limited by the effects of using a finite-size aperture and by imperfect alignment of the light from the light source and the control laser \citep{Learner88}. To obtain a wavenumber scale which is accurate to better than 1 part in $10^{7}$, a multiplicative correction is applied using a correction factor, $k_{\mathrm{eff}}$, such that
\begin{equation}
  \sigma_{\mathrm{corr}}=(1+k_{\mathrm{eff}})\sigma_{\mathrm{obs}}\,,
  \label{correq}
\end{equation}
where $\sigma_{\mathrm{corr}}$ is the corrected wavenumber and $\sigma_{\mathrm{obs}}$ is the observed, uncorrected, wavenumber.

The factor $k_{\mathrm{eff}}$ is accurately determined by measuring positions of one or several well-known internal wavenumber standard lines. In principle, it is possible to use only one calibration line \citep{Salit04}, but to reduce the uncertainty of the calibration, several calibration lines have been used for each recorded spectrum. The use of internal calibration lines, in this case Ar{\sc \,ii} lines from the carrier gas in the HCL, helps to ensure, to as high degree as possible, that the light from the species being used for wavelength calibration illuminates the entrance aperture of the FT spectrometer in the same way as that of the species being investigated.

\begin{figure}
 \includegraphics[width=84mm]{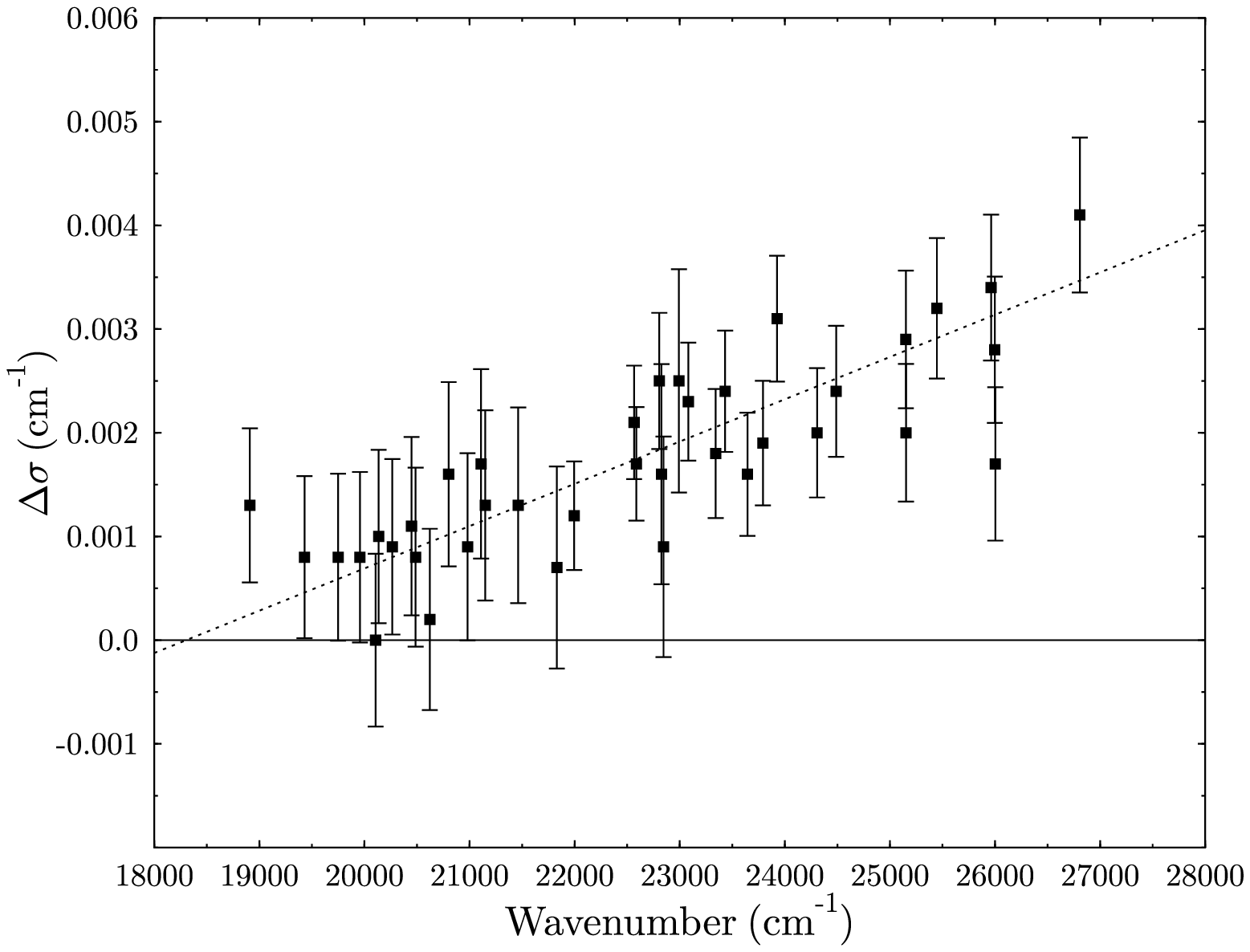}
 \caption{Comparison between 4s-4p Ar{\sc \,ii} wavenumbers from \citet{Whaling95} and \citet{Norlen73} ($\Delta \sigma = \sigma_{\mathrm{Whaling}}-\sigma_{\mathrm{Norl\acute{e}n}}$). The error bars include the quoted uncertainties of both Whaling's and Norl\'en's values.}
 \label{Whalingfig}
\end{figure}

Ar{\sc \,ii} lines are commonly used for calibrating spectra and there are in principle two sets of standard lines available. The general opinion is now that the work by \citet{Whaling95} should be used instead of the work by \citet{Norlen73}, which has previously been the standard. The wavenumbers of Whaling are stated to have a higher absolute accuracy than Norl\'en's and are measured using FT spectrometry with molecular CO lines as wavenumber standards. There is an obvious difference between the two sets of lines, which increases with increasing wavenumber, see Fig.~\ref{Whalingfig} where the differences in 4s-4p transitions are shown. Whaling's wavenumbers are higher by 0.001 to 0.003\,cm$^{-1}$ in the wavenumber range 21000 to 26000\,cm$^{-1}$ where recording number I is calibrated. More quantitative comparisons between the two sets of measurements have been made by \citet{Whaling95,Whaling02} and \citet{Nave04}.

\begin{figure}
 \includegraphics[width=84mm]{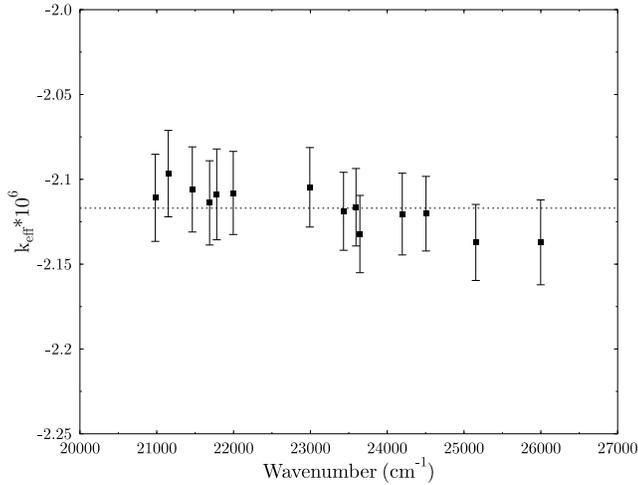}
 \caption{Calibration of recording number I using 4s-4p Ar{\sc \,ii} lines from \citet{Whaling95}.}
 \label{CalibIfig}
\end{figure}

The calibration lines selected for recording number I are 14 Ar{\sc \,ii} 4s-4p transitions from \citet{Whaling95} with signal-to-noise ratios larger than 100. Most of the lines used are transitions recommended by \citet{Learner88} to be least affected by pressure shifts. The pressure shift generally increases with excitation. Due to the few low levels in Ar{\sc \,ii} this leads to the fact that excitation and thus pressure shift increases with the wavenumber of the lines. The pressure in the cathode used by \citet{Whaling95} (1--4\,torr) differs from the pressure used in this work (0.9--1.0\,torr). Using the least pressure sensitive 4s-4p transitions minimizes the possible effects from this pressure difference. Fig.~\ref{CalibIfig} shows a plot of $k_{\mathrm{eff}}$ versus $\sigma$ for the 14 Ar{\sc \,ii} calibration lines in recording number I. The plot deviates slightly from the expected horizontal straight line even with the restricted set of 4s-4p lines. It appears to change by 3 parts in $10^8$ over 5000\,cm$^{-1}$ giving rise to an uncertainty in the calibration. The adopted correction factor is derived by taking the weighted mean of the correction factors from the 14 calibration lines, where the weight is scaled with the inverse variance of each correction factor. The error bars in Fig.~\ref{CalibIfig} include both Whaling's quoted uncertainty and the uncertainty in our line fitting procedure, see Sections~\ref{fitting} and~\ref{uncertainties}. The dashed line in Fig.~\ref{CalibIfig} marks the adopted correction factor.

\begin{figure}
 \includegraphics[width=84mm]{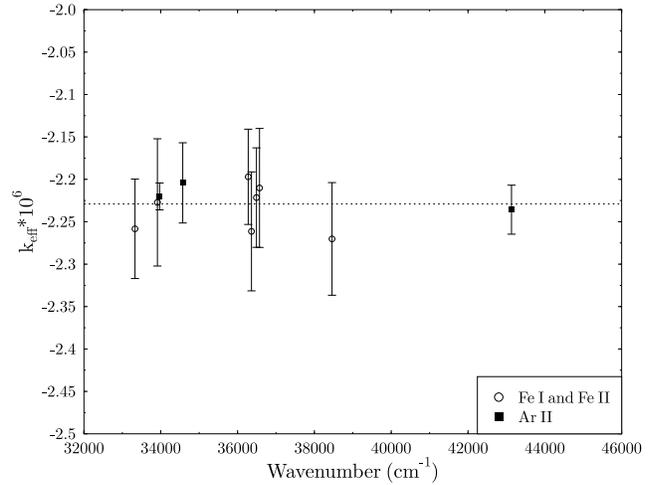}
 \caption{Calibration of recording number II using Fe{\sc \,i} and Fe{\sc \,ii} lines, from the overlap region, together with 4s-4p Ar{\sc \,ii} lines from \citet{Whaling95}.}
 \label{CalibIIfigFe}
\end{figure}

In recordings number II and III, covering higher wavenumbers, there are an insufficient number of reliable Ar{\sc \,ii} calibration lines available in our spectra. Only three 4s-4p transitions have sufficient signal-to-noise-ratios ($>100$) for calibration purposes and these might possibly be affected by pressure shifts due to their higher wavenumbers (see previous discussion). A transfer of calibration between the two wavenumber regions was therefore used. In recording number I, which includes the 14 Ar{\sc \,ii} calibration lines, a number of Fe{\sc \,i} and Fe{\sc \,ii} lines were measured and calibrated. These lines are located in the overlap of the two wavenumber intervals and are observed in all three recordings. They could therefore be used together with three Ar{\sc \,ii} lines for calibration in recordings number II and III. The uncertainty of the calibration of the higher wavenumbers are slightly larger due to this two-step calibration. Fig.~\ref{CalibIIfigFe} shows a plot of $k_{\mathrm{eff}}$ versus $\sigma$ for the Fe{\sc \,i}, Fe{\sc \,ii} and Ar{\sc \,ii} calibration lines used in recording number II. The Fe lines and Ar{\sc \,ii} lines seem to give the same correction factor implying that the possible pressure shift in the Ar{\sc \,ii} lines used from \citet{Whaling95} is negligible.

\begin{figure}
 \includegraphics[width=84mm]{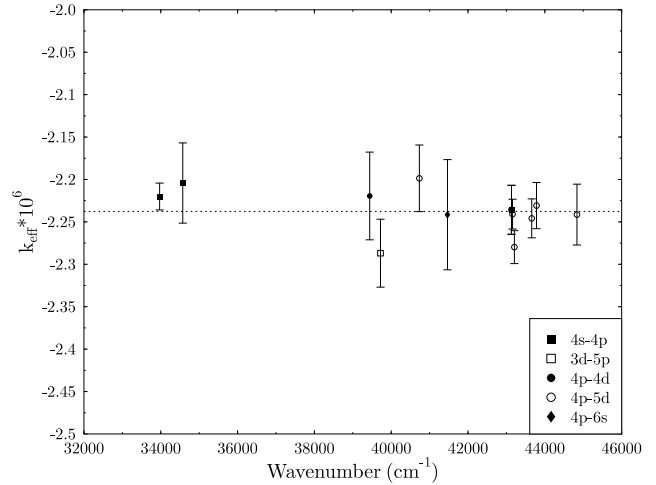}
 \caption{Calibration of recording number II using Ar{\sc \,ii} lines from \citet{Whaling95}. Besides 4s-4p transitions, this calibration includes strong 3d-5p, 4p-4d, 4p-5d and 4p-6s transitions.}
 \label{CalibIIfigAr}
\end{figure}

To investigate the possible shifts of the Ar{\sc \,ii} lines further another calibration of recording number II was made by including strong 4s-4p, 3d-5p, 4p-4d, 4p-5d and 4p-6s Ar{\sc \,ii} transitions from \citet{Whaling95}, see Fig.~\ref{CalibIIfigAr}. The resulting correction factor differs no more than 1 part in $10^8$ from the value derived from the Fe calibration lines in the overlap region, which is well within the estimated uncertainty.

The HCL is only stable within a small pressure range, but to investigate possible effects of pressure shifts measurements were made with three different pressures (0.8, 1.0 and 1.2\,torr) in the light source, keeping the rest of the parameters (i.e. the current and the quantity of metal in the HCL, and the spectrometer settings) constant. The possible effects of shifting wavenumbers and varying calibration factors, for this small pressure range, were well within the uncertainties.

\subsection{Line fitting}\label{fitting}

\begin{figure*}
 \includegraphics{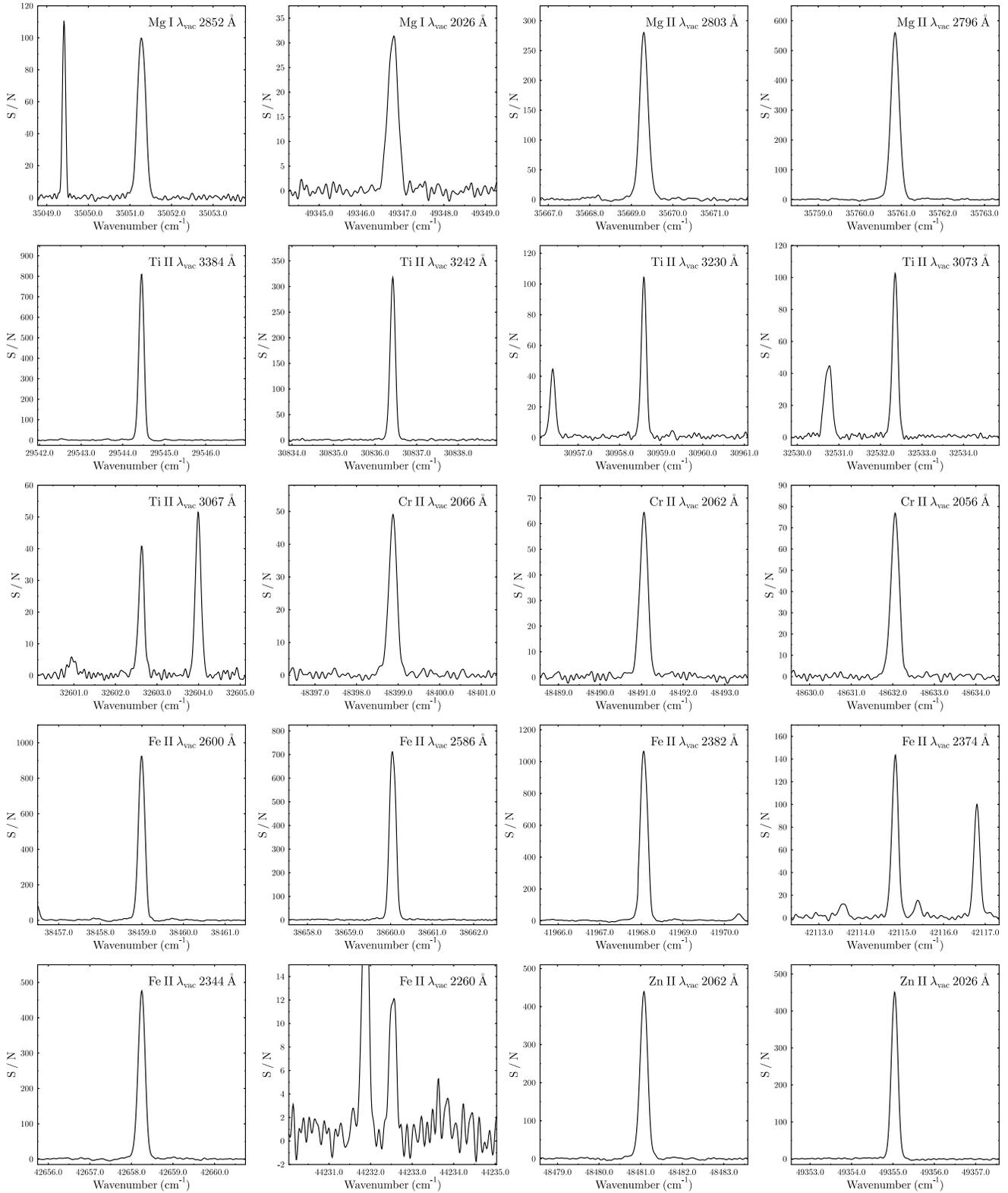}
 \caption{The observed profiles of the Mg, Ti, Cr, Fe and Zn lines in this study. The lines are centred in a wavenumber interval of $\pm$2.5\,cm$^{-1}$.}
 \label{20linesfig}
\end{figure*}

All lines except the Mn lines show no structure and consist of an apparently symmetric peak profile, see Fig.~\ref{20linesfig} where the observed profiles of the Mg, Ti, Cr, Fe and Zn lines are plotted. Possible unresolved isotope shifts are discussed below in Section~\ref{isotope}. The symmetric lines were fitted with Voigt profiles using a least-square procedure included in the FT spectrometry analysis program {\sc Xgremlin} \citep{Nave97}, which is based on the {\sc Gremlin} code \citep{Brault89}.

\subsection{Mn hyperfine structure}\label{mnhyperfine}
\begin{figure*}
 \includegraphics{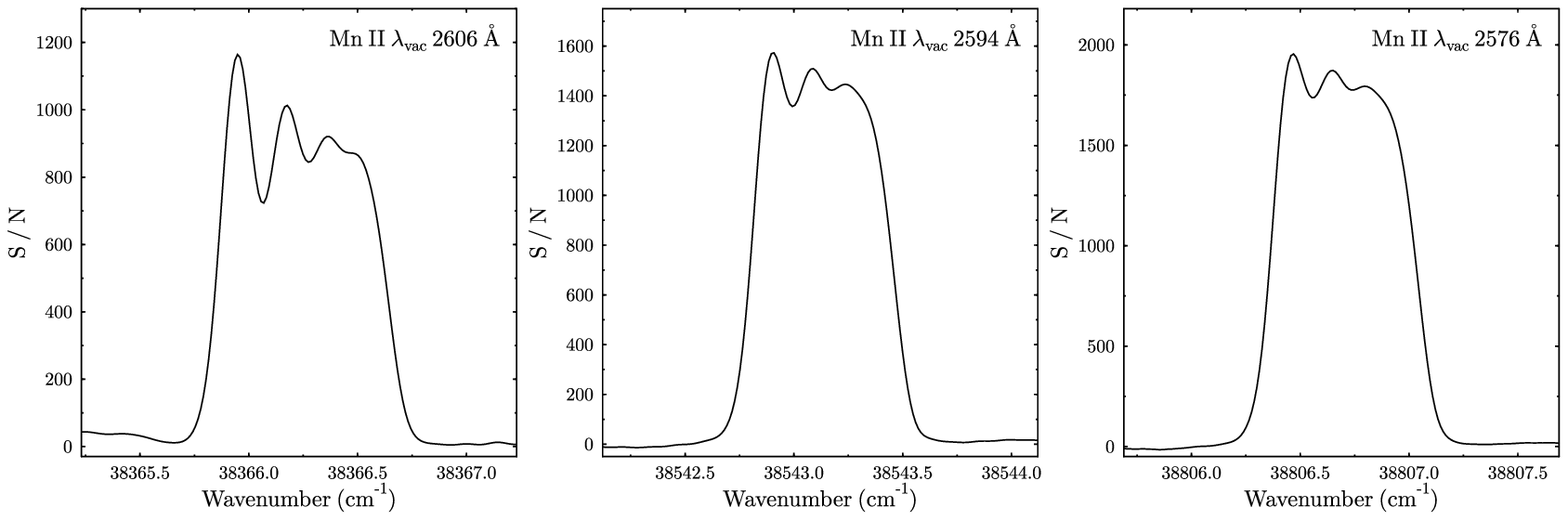}
 \caption{The observed profiles of the Mn{\sc \,ii} lines in this study, showing unresolved hyperfine structure. The lines are centred in a wavenumber interval of $\pm$1.0\,cm$^{-1}$.}
 \label{Mnstructurefig}
\end{figure*}

The manganese lines consist of several peaks arising from the hyperfine structure. Mn consists of only one natural isotope, $^{55}\mathrm{Mn}$, which has a nuclear spin of $\mathrm{I}=5/2$ and a magnetic dipole moment of 3.45. This gives rise to a large hyperfine structure in the lines. The structure is however only partially resolved, see Fig.~\ref{Mnstructurefig}. The structure of Mn{\sc \,ii} resonance lines is previously reported \citep{Kling00}, but the hyperfine constants for the energy levels included in the present investigation have not been measured until recently \citep{Blackwell05}.

Since the structure of the Mn{\sc \,ii} lines is unresolved the wavenumbers were determined by measuring the Centre-of-Gravity (CoG) wavenumbers of the lines instead of fitting a Voigt profile as for the other lines. The CoG of a line was determined by choosing a start and an end point symmetrically located at each side of the line profile. The area and the CoG were then calculated for the region between the two points. This was done 20 times for each line, over regions with varying widths, and a mean was adopted as the wavenumber of the line. Due to the strength of the lines this gave a consistent result with an estimated uncertainty small enough to match the required precision for many-multiplet analysis, see Section~\ref{uncertainties} and~\ref{introduction}.

\begin{figure}
 \includegraphics[width=84mm]{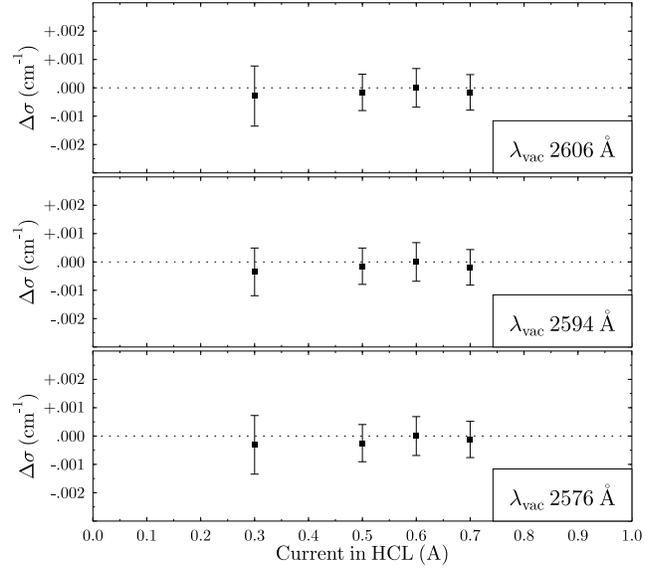}
 \caption{Test for optical thickness for the Mn{\sc \,ii} lines. The error bars represent the relative uncertainty of the Centre-of-Gravity for the Mn{\sc \,ii} lines, not including the uncertainty of the calibration lines.}
 \label{Mnopticalfig}
\end{figure}

The effect of a possible self-absorption of the Mn{\sc \,ii} lines could be that the observed CoG is shifted towards the weaker hyperfine components. To test this recordings were made with different currents through the HCL. This would lead to different plasma densities in the cathode and thus different amount of self-absorption. The lower the current, the lower the density and self-absorption. Fig.~\ref{Mnopticalfig} shows the differences in the measured CoG of the Mn{\sc \,ii} lines at different currents. No apparent tendency is noticed and the shifts of the wavenumbers are well within the uncertainties.

When applying the measured hyperfine constants by \citet{Blackwell05} and fitting the observed profiles with all hyperfine components, the resulting CoG wavenumbers agreed well, within the uncertainties, with our unfitted CoG wavenumbers.

\begin{figure}
 \includegraphics[width=84mm]{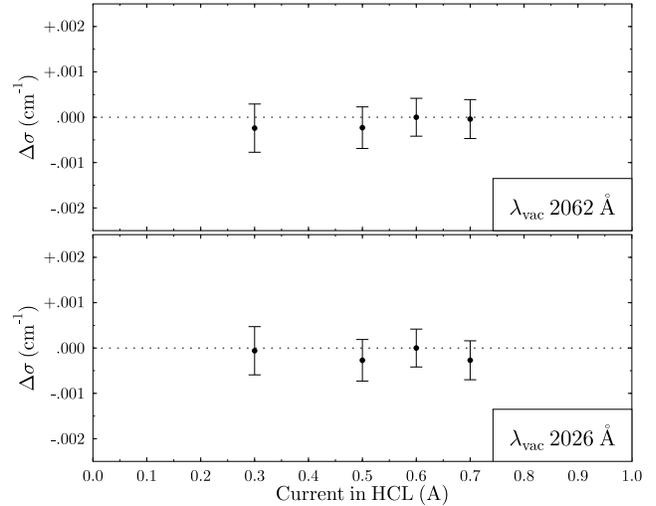}
 \caption{Test for optical thickness for the Zn{\sc \,ii} lines. The error bars represent the relative uncertainty of the wavenumber for the Zn{\sc \,ii} lines, not including the uncertainty of the calibration lines.}
 \label{Znopticalfig}
\end{figure}

\subsection{Possible isotope structure}\label{isotope}
Although the lines from Mg, Ti, Cr, Fe and Zn appear to be symmetric and the residuals from the line fitting showed no asymmetry, a further investigation of possible unresolved isotope structure was carried out. For Cr and Fe any isotope or hyperfine structure should be negligible compared to the Doppler width and they both have one dominating isotope ($^{52}\mathrm{Cr}$:~84 and $^{56}\mathrm{Fe}$:\,92 per~cent). Since the Fe{\sc \,ii} lines are of great importance to the MM method, the observed line profiles were compared to synthetic profiles consisting of components from the most abundant isotopes ($^{54}\mathrm{Fe}$:\,6, $^{56}\mathrm{Fe}$:\,92 and $^{57}\mathrm{Fe}$:\,2 per~cent). Transition shifts larger than $\sim\,0.06 {\rm \,cm}^{-1}$, between the $^{54}\mathrm{Fe}$ and $^{56}\mathrm{Fe}$ isotopes, violated the observed profile and the shifts should therefore be smaller than that.

Both Ti and Mg consist of one even isotope each with a large relative abundance and a couple of other isotopes with relative abundances between 5 and 10 per~cent. The lines measured showed no form of self-absorption, which could have been a problem since unresolved structure might cause a small shift in the wavenumbers of the lines had they been affected by self-absorption. A thorough investigation of the isotope structure of the Mg{\sc \,i} $\lambda$2852\,\AA\ and Mg{\sc \,ii} $\lambda\lambda$2796,2803\,\AA\ lines has been made by \citet{Pick98}.

The Zn{\sc \,ii} doublet was investigated more thoroughly since Zn has three main isotopes ($^{64}\mathrm{Zn}$:\,49, $^{66}\mathrm{Zn}$:\,28 and $^{68}\mathrm{Zn}$:\,19 per~cent) and the lines showed clearly self-absorbed profiles during the test runs. To ensure that the possible self-absorption did not shift the lines the discharge current was varied between 0.3 and 0.7\,A. This would produce different plasma densities and thus different amount of self-absorption. In Fig.~\ref{Znopticalfig} the differences in the fitted wavenumbers of the Zn{\sc \,ii} lines are plotted at different discharge currents. Similar to the Mn{\sc \,ii} lines, no shift or tendency is apparent. The transition isotope shifts for the Zn{\sc \,ii} $\lambda$2026\,\AA\ line have been measured, using laser technique, by \citet{Matsubara03} reporting shifts of 0.670--0.676\,GHz (corresponding to 0.022--0.023\,cm$^{-1}$) for the mentioned isotopes. This is small compared to the Doppler width of about 0.2\,cm$^{-1}$ in our observed lines.


\section{Results}\label{results}
The measured wavenumbers, wavelengths and uncertainties of the 23 lines are presented in Table~\ref{resultstable}. The wavenumbers are compared to previously published values.

\begin{table*}
\caption{Results. Wavelengths, wavenumbers and comparison to previous values. $\lambda\mathrm{_{air}}$ is derived by the \citet{Edlen66} dispersion formula for air at 15$^{\circ}$C and atmospheric pressure.}
 \begin{tabular}{lccccll}
 \hline
 Species & $\lambda\mathrm{_{air}}$ & $\lambda\mathrm{_{vac}}$ & Wavenumber & Recording & Previous Value & Reference\\
 {} & (\AA ) & (\AA ) & (cm$^{-1}$) & number & (cm$^{-1}$) & {}\\
 \hline
 Mg{\sc \,i} & 2852.1248 & $2852.9628\pm0.0001$ & $35051.280\pm0.002$ & I & $35051.277\pm0.001$ & \citet{Pick98}\\
 Mg{\sc \,i} & 2025.8224 & $2026.4750\pm0.0001$ & $49346.771\pm0.003$ & II & $49346.729\pm0.070$ & \citet{Risberg65}\\
 Mg{\sc \,ii} & 2802.7052 & $2803.5311\pm0.0001$ & $35669.303\pm0.002$ & II & $35669.298\pm0.002$ & \citet{Pick98}\\
 Mg{\sc \,ii} & 2795.5299 & $2796.3540\pm0.0001$ & $35760.851\pm0.002$ & II & $35760.848\pm0.002$ & \citet{Pick98}\\
 Ti{\sc \,ii} & 3383.7584 & $3384.7300\pm0.0002$ & $29544.454\pm0.001$ & I & $29544.451\pm0.003$ & Zapadlik et al. (2001)$^{a}$\\
 Ti{\sc \,ii} & 3241.9825 & $3242.9180\pm0.0002$ & $30836.426\pm0.001$ & I & $30836.422\pm0.003$ & Zapadlik et al. (2001)$^{a}$\\
 Ti{\sc \,ii} & 3229.1895 & $3230.1217\pm0.0002$ & $30958.586\pm0.001$ & I & $30958.582\pm0.003$ & Zapadlik et al. (2001)$^{a}$\\
 Ti{\sc \,ii} & 3072.9701 & $3073.8629\pm0.0001$ & $32532.355\pm0.001$ & I & $32532.351\pm0.003$ & Zapadlik et al. (2001)$^{a}$\\
 Ti{\sc \,ii} & 3066.3463 & $3067.2375\pm0.0002$ & $32602.627\pm0.002$ & I & $32602.623\pm0.003$ & Zapadlik et al. (2001)$^{a}$\\
 Cr{\sc \,ii} & 2065.5039 & $2066.1639\pm0.0001$ & $48398.871\pm0.002$ & II & $48398.868\pm0.002$ & \citet{Pick00}\\
 Cr{\sc \,ii} & 2061.5767 & $2062.2359\pm0.0001$ & $48491.057\pm0.002$ & II & $48491.053\pm0.002$ & \citet{Pick00}\\
 Cr{\sc \,ii} & 2055.5987 & $2056.2568\pm0.0001$ & $48632.058\pm0.002$ & II & $48632.055\pm0.002$ & \citet{Pick00}\\
 Mn{\sc \,ii} & 2605.6804 & $2606.4588\pm0.0002$ & $38366.230\pm0.003$ & II & $38366.227\pm0.003$ & Blackwell-W. et al. (2005)\\
 Mn{\sc \,ii} & 2593.7211 & $2594.4967\pm0.0002$ & $38543.121\pm0.003$ & II & $38543.122\pm0.003$ & Blackwell-W. et al. (2005)\\
 Mn{\sc \,ii} & 2576.1039 & $2576.8753\pm0.0002$ & $38806.689\pm0.003$ & II & $38806.689\pm0.003$ & Blackwell-W. et al. (2005)\\
 Fe{\sc \,ii} & 2599.3953 & $2600.1722\pm0.0001$ & $38458.991\pm0.002$ & II & $38458.987\pm0.002$ & \citet{Nave91}\\
 Fe{\sc \,ii} & 2585.8756 & $2586.6494\pm0.0001$ & $38660.052\pm0.002$ & III & $38660.049\pm0.002$ & \citet{Nave91}\\
 Fe{\sc \,ii} & 2382.0375 & $2382.7641\pm0.0001$ & $41968.065\pm0.002$ & II & $41968.064\pm0.002$ & \citet{Nave91}\\
 Fe{\sc \,ii} & 2373.7354 & $2374.4601\pm0.0001$ & $42114.836\pm0.002$ & II & $42114.833\pm0.002$ & \citet{Nave91}\\
 Fe{\sc \,ii} & 2343.4949 & $2344.2128\pm0.0001$ & $42658.243\pm0.002$ & II & $42658.240\pm0.002$ & \citet{Nave91}\\
 Fe{\sc \,ii} & 2260.0797 & $2260.7793\pm0.0003$ & $44232.534\pm0.006$ & II & $44232.512\pm0.020$ & \citet{Johansson78}\\
 Zn{\sc \,ii} & 2062.0010 & $2062.6603\pm0.0001$ & $48481.081\pm0.002$ & II & $48481.077\pm0.002$ & \citet{Pick00}\\
 Zn{\sc \,ii} & 2025.4844 & $2026.1369\pm0.0001$ & $49355.005\pm0.002$ & II & $49355.002\pm0.002$ & \citet{Pick00}\\
 \hline
 \end{tabular}
 \smallskip
 \flushleft
\hspace{5mm}$^{a}$Zapadlik, I., Johansson, S. and Litz\'en, U., Private communication 2006. Published in the compilation by \citet{Morton03}
 \label{resultstable}
\end{table*}

For the many-multiplet method it is the relative shifts of the lines that is most important. The present investigation therefore includes lines from different species measured in the same recordings, which reduces the uncertainties of the relative wavenumbers. Previous FT spectrometry measurements of the Cr{\sc \,ii}, Zn{\sc \,ii} \citep{Pick00}, Mg{\sc \,ii} lines and the Mg{\sc \,i} $\lambda$2852\,\AA\ line \citep{Pick98} have been made for the same reason as this investigation \citep[these are included in][]{Morton03}. \citet{Pick00} measured the lines at two different laboratories (Imperial College, London, and Lund) to ensure a minimization of systematic errors. At Lund a composite cathode was used, similar to the present investigation, to reduce the relative uncertainties. The published values for Cr and Zn were, however, weighted means of the ICL and LU results.

\begin{figure}
 \includegraphics[width=84mm]{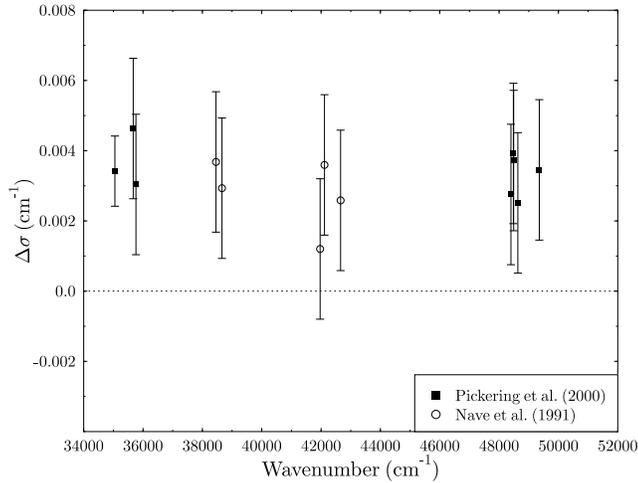}
 \caption{Comparison to previous measurements ($\Delta \sigma = \sigma_{\mathrm{this work}}-\sigma_{\mathrm{previous}}$) calibrated with Ar{\sc \,ii} lines from \citet{Norlen73}. The error bars represent the stated uncertainties of the previous values.}
 \label{Comparisonfig}
\end{figure}

One significant difference between the previous Lund-ICL and the present investigations is the calibration. The previous wavenumbers have been calibrated directly or indirectly with Ar{\sc \,ii} lines from \citet{Norlen73}. There is a systematic difference between these wavenumbers compared to the \citet{Whaling95} wavenumbers used for calibration in this work, see Section~\ref{calibration}. In Fig.~\ref{Comparisonfig} the difference between the wavenumbers from this work and the previous published wavenumbers are plotted showing an offset of about 0.003\,cm$^{-1}$ stemming from the difference in the calibration lines used. Included in the plot are wavenumbers of Mg{\sc \,i}, Mg{\sc \,ii}, Cr{\sc \,ii}, Zn{\sc \,ii} \citep{Pick00} and Fe{\sc \,ii} \citep{Nave91} lines, which all are measured using Fourier Transform spectrometers. 

The difference between the Whaling and Norl\'en Ar{\sc \,ii} calibration lines have been discussed by \citet{Nave04}, where a correction is recommended for the previously published Fe{\sc \,ii} lines in \citet{Nave91}. The previous wavenumbers are recommended to be increased by $6.7$ parts in $10^{8}$. For the 5 Fe{\sc \,ii} lines studied in this investigation the correction would imply that the wavenumbers from \citet{Nave91} should be increased with about $0.003$ cm$^{-1}$. This confirms that the offset in Fig.~\ref{Comparisonfig} is mainly stemming from the difference in the sets of calibration lines.

Recently also FT spectrometry measurements of the wavelengths for the Mn{\sc \,ii} lines have been published, together with hyperfine constants for the levels included, by \citet{Blackwell05}. Also these wavenumbers were calibrated with Ar{\sc \,ii} lines from \citet{Norlen73} and should therefore be lower, by 0.003 cm$^{-1}$, than the wavenumbers measured in this investigation. The estimated uncertainties are however larger than for symmetric lines, both for our measurements and for those of Blackwell-Whitehead, due to the hyperfine structure. The measurements agree, within the uncertainties, when the calibration difference is taken into account.

The previous values for Ti{\sc \,ii} are unpublished FT spectrometry measurements by Zapadlik et al. \citep[private communication 2006 and published in][]{Morton03}, also calibrated with Ar{\sc \,ii} lines from \citet{Norlen73}, while the Mg{\sc \,i} $\lambda$2026\,\AA\ and Fe{\sc \,ii} $\lambda$2260\,\AA\ lines are measured using grating technique and have much lower precision and calibration accuracy.

\subsection{Uncertainties}\label{uncertainties}
The absolute wavenumber uncertainty contains two main factors: the uncertainty of the calibration and the uncertainty of the determination of the line position.
The standard deviation $s(\sigma)$ of the line position $\sigma$ of a fitted line profile can be written as \citep{Brault87,Sikstrom02}
\begin{equation}\label{unceq}
s(\sigma)=\alpha_{\sigma}\frac{\sqrt{\mathrm{d}x}}{S}\sqrt{w}=\alpha_{\sigma}\frac{w}{S\sqrt{n}}\,,
\end{equation}
where $S$ is the signal-to-noise ratio for the line, $\mathrm{d}x$ is the resolution interval and $w$ is the FWHM of the line. The number of points across the width of the line is given by $n=w/\mathrm{d}x$. $\alpha_{\sigma}$ is a numerical constant depending on the shape of the line. For a pure Gaussian profile the value is $\alpha_{\sigma,G}=0.69\pm0.02$ and for a pure Lorentzian profile the value is $\alpha_{\sigma,L}=0.80\pm0.13$ \citep{Sikstrom02}.

The uncertainty from the calibration is derived from the uncertainties stated by \citet{Whaling95}, from the uncertainties of the line positions of the calibration lines and from the standard deviation of the calibration factors.

For the Mn{\sc \,ii} lines the CoG is given and the uncertainty is estimated by calculating the standard deviation of 20 different measurements of the CoG (in one recording) of each line, adding the standard deviation given by making the same measurements in different recordings (with different lamp conditions) and finally including the uncertainty of the calibration. This resulted in a total uncertainty of $\pm 0.003$\,cm$^{-1}$ for the three Mn{\sc \,ii} lines.

The difference in wavenumbers between two lines measured in the same recording follows from Eq.~\ref{correq} and can be expressed as
\begin{equation}
\sigma_{1}-\sigma_{2}=(1+k_{\mathrm{eff}})(\sigma_{obs,1}-\sigma_{obs,2}) \,,
\end{equation}
where $k_{\mathrm{eff}}$ is the calibration factor for the recorded spectrum.
By applying the law of propagation of uncertainty, the expression for the uncertainty of the difference in wavenumber becomes 
\begin{eqnarray}
s\left( \sigma_{1}-\sigma_{2}\right) = \left\{ \left( 1+k_{\mathrm{eff}}\right)^{2}\left[ s(\sigma_{obs,1})^2 + s(\sigma_{obs,2})^2\right] \right. \nonumber \\
\left. + \left( \sigma_{obs,1}-\sigma_{obs,2}\right)^2 s(k_{\mathrm{eff}})^2 \right\} ^{1/2} \,,
\end{eqnarray}
where $s(\sigma_{obs,i})$ is the uncertainty of the determination of the line position for line $i$, see Eq.~\ref{unceq}, and $s(k_{\mathrm{eff}})$ is the uncertainty of the calibration factor for the observed spectrum, see Table~\ref{exptable}. This means that for symmetric lines with high signal-to-noise ratios ($>$250) the relative wavenumbers (for lines within the same spectral recording) are determined with an uncertainty of $\pm 0.0005$\,cm$^{-1}$, or better. The uncertainty is smallest for lines which are close in wavenumber, due to the linear correction, and then depends mostly on the line fitting.


\section{Conclusions}\label{conclusions}

Constraints on possible cosmological variations in the fine-structure constant rely crucially on comparison with accurate laboratory standards. Here we have measured 23 transitions, with particular emphasis on {\it relative} wavelength precision, which are central to many-multiplet (MM) analyses of QSO absorption spectra. The precision is improved by measuring the lines simultaneously from a composite light source. The lines are calibrated against new secondary standard lines of Ar{\sc \,ii}, produced in the same light source. The estimated absolute uncertainty is 0.1\,m\AA\ for 15 lines, 0.2\,m\AA\ for 7 lines and 0.3\,m\AA\ for one line. It should be pointed out that the Mn{\sc \,ii} lines are severely broadened due to unresolved hyperfine structure, and that their wavelengths represent centre-of-gravity values. The relative wavenumbers for strong, symmetric lines in the same spectral recording are determined with an uncertainty of 0.03\,m\AA\ at 2500\,\AA.

This is the first time the 5 transitions of Ti{\sc \,ii} and the Mg{\sc \,i} $\lambda$2026 and Fe{\sc \,ii} $\lambda$2260 transitions have been measured with sufficient precision for use in the MM method. The remaining 16 transitions of Mg{\sc \,i}, Mg{\sc \,ii}, Cr{\sc \,ii}, Mn{\sc \,ii}, Fe{\sc \,ii} and Zn{\sc \,ii} have been measured to as high precision before using a different set of calibration lines. {\it The importance of repeated laboratory measurements cannot be emphasised enough given the role they play in fundamental physics experiments at cosmological distances and look-back times}. 

Due to the newly adopted wavenumbers for a set of Ar II calibration lines the absolute wavenumbers of the majority of the lines have been shifted by about 0.003\,cm$^{-1}$ (about 0.1-0.2\,m\AA) compared to previous values \citep[see e.g.][]{Morton03}. The wavelengths which have been measured to as high a precision before agree with the present wavelengths to within the uncertainties, when this calibration difference is taken into account. This leads to a greater confidence in these wavelengths, while the high relative precision in this work reduces the potential for systematic effects in the MM method. As all 23 lines measured are ground state transitions they also usually appear as sharp interstellar lines in stellar spectra.

\section*{Acknowledgments}
We are grateful to Dr.~U.~Litz\'{e}n for valuable comments on the manuscript. This project is supported by a grant (SJ) from the Swedish Research Council. MTM thanks PPARC for an Advanced Fellowship.



\vspace{0.5cm}\small\noindent This paper
has been typeset from a \TeX / \LaTeX\ file prepared by the author.

\label{lastpage}

\end{document}